\documentclass[cpp,fleqn]{w-art}
\usepackage{times}
\usepackage{w-thm}
\usepackage[]{graphicx}
\begin{document}
\DOIsuffix{theDOIsuffix}
\Volume{42}
\Issue{1}
\Month{01}
\Year{2003}
\pagespan{3}{}
\Receiveddate{}
\Reviseddate{}
\Accepteddate{}
\Dateposted{}



\title[Strong-coupling theory for a polarizable planar colloid]{Strong-coupling 
theory for a polarizable planar colloid}


\author[L. \v{S}amaj]{Ladislav \v{S}amaj\inst{1,2}} 
\address[\inst{1}]{Institute of Physics, Slovak Academy of Sciences, 
D\'ubravsk\'a cesta 9, 84511 Bratislava, Slovakia}
\author[E. Trizac]{Emmanuel Trizac\inst{2}}
\address[\inst{2}]{Laboratoire de Physique Th\'eorique et Mod\`eles Statistiques, 
UMR CNRS 8626, Universit\'e Paris-Sud, 91405 Orsay, France}
\begin{abstract}
We propose a strong-coupling analysis of a polarizable planar interface,
in the spirit of a recently introduced Wigner-Crystal formulation. 
The system is made up of two moieties: a semi-infinite medium 
($z<0$) with permittivity
$\varepsilon'$ while the other half space in $z>0$ is occupied by a
solution with permittivity $\varepsilon$, and mobile counter-ions
(no added electrolyte). The interface at $z=0$ bears a uniform surface charge.
The counter-ion density profile is worked 
out explicitly for both repulsive and attractive dielectric image 
cases.  
\end{abstract}
\maketitle                   





{\bf Introduction}.
Strong coulombic couplings can conveniently be realized in colloidal systems
by increasing the valence of micro-ions \cite{Grosberg02,Levin02,Naji05}. The
counter-intuitive effects such as like-charge attraction or overcharging 
that ensue \cite{Grosberg02,Levin02,Naji05}, have been mostly studied under the
simplifying assumption that the dielectric permittivity of colloidal
particles coincides with that of the surrounding solvent --assumed to
be a structure-less medium, characterized by its sole static dielectric response.
However, realistic systems, of biological interest in particular,
often exhibit highly inhomogeneous dielectric structure, with polarizable colloids
having a static dielectric constant ($\varepsilon' \leq 10$), much smaller
than that of the solvent ($\varepsilon \simeq 80$ for water). It is therefore desirable to
include colloidal polarizability into existing strong-coupling treatments
(see below for a precise definition of the relevant coupling parameter).
This is the purpose of the present contribution.

\begin{figure}[htb]
\sidecaption
\includegraphics[width=0.3\textwidth,clip]{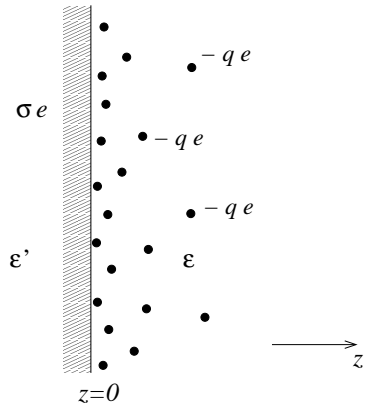}
\caption{The one-plate geometry considered.
Counter-ions of charge $-q e$, $e$ being the electron charge, 
are immersed in a solvent of dielectric
constant $\varepsilon$ in the region $z>0$ while the wall, a large colloid
with surface charge $\sigma e$ occupying the $z<0$ half-space,
has the dielectric constant $\varepsilon'$. From these quantities and the inverse 
temperature $\beta$, it is customary to define the coupling parameter 
$\Xi=2 \pi q^3 \ell_B^2 \sigma$, where $\ell_B=\beta e^2 /\varepsilon$ is the 
so-called Bjerrum length. When $\Xi \ll 1$, mean-field treatments
are legitimate, while the strong-coupling regime of interest here corresponds to the limit
of large $\Xi$.}
\label{fig:1} 
\end{figure}

To our knowledge, the dielectric inhomogeneities were studied in the 
leading strong-coupling order only for the geometry of two parallel plates 
\cite{Kanduc07} 
where  counter-ions are confined in a slab,
and similarly in Ref. \cite{BdSL11}, for a spherical macro-ion
in a concentric spherical Wigner-Seitz cell.
It was realized that while the image charge effects are generally small in 
the weak-coupling regime, they become relevant in the strong-coupling limit.
Our goal here is to obtain exact results for a single polarizable planar 
colloid (hatched region in Fig. \ref{fig:1}), in the strong-coupling limit
(see Fig. \ref{fig:1} for a definition of the coupling parameter 
$\Xi$). To this end, the Wigner-Crystal formulation of Strong-Coupling (SC) 
theory, recently introduced \cite{Samaj11a}, is modified to account 
for dielectric image charges. 
We emphasize that this approach has been shown to be in excellent agreement 
with Monte Carlo simulations, in contrast to previous fugacity SC expansions
\cite{Naji05}.
The idea is to consider the first relevant
large $\Xi$ excitations around the ground state structure, that is a Wigner
Crystal realized when, strictly speaking, $\Xi\to \infty$ \cite{Samaj11a}.

We define the dielectric jump between the two moieties as 
$\Delta = (\varepsilon-\varepsilon')/(\varepsilon+\varepsilon')$.
The image of a charge $e$ at position ${\bf r}=(x,y,z>0)$ has the charge of 
strength $e^*= e \Delta$ and is localized at position ${\bf r}^*=(x,y,-z<0)$.
Counter-ions interact via the Coulomb interaction potential 
$u({\bf r},{\bf r}') = u_0({\bf r},{\bf r}') + u_{\rm im}({\bf r},{\bf r}') , 
$
where $u_0({\bf r},{\bf r}')$ and $u_{\rm im}({\bf r},{\bf r}')$
are the direct and image Coulomb potentials, respectively, 
defined by
\begin{equation}
u_0({\bf r},{\bf r}') = \frac{1}{\varepsilon\vert{\bf r}-{\bf r}'\vert} , \quad 
u_{\rm im}({\bf r},{\bf r}') = 
\frac{\Delta}{\varepsilon\vert{\bf r}^*-{\bf r}'\vert} . 
\end{equation}
The total interaction energy of the counter-ions at positions 
$\{ {\bf r}_j\}_{j=1}^N$ in the half-space $z>0$ reads
\begin{equation} \label{eq:energypi}
E(\{ {\bf r}_j\}) =  \sum_{j=1}^N 
\left[ \frac{2\pi qe^2\sigma}{\varepsilon} (1+\Delta) z_j
+ \frac{(qe)^2}{4\varepsilon} \frac{\Delta}{z_j} \right] 
+ \frac{(qe)^2}{2} \sum_{j,k=1\atop (j\ne k)}^N u({\bf r}_j,{\bf r}_k) .
\end{equation}
Here, the two one-body terms in the square brackets originate in 
the interaction of the particle with the surface charge plus its image 
(therefore the factor $1+\Delta$) and in the interaction of the particle 
with its own charge image.
We shall study separately the cases of attractive $(\Delta<0)$
and repulsive $(\Delta>0)$ image charges.
In addition, we shall consider dimensionless quantities: 
$\tilde z = z /\mu$ where $\mu=1/(2 \pi q \ell_B \sigma)$ is the Gouy-Chapman
length, and likewise, counter-ion densities will 
be made unit-less by the transformation $\tilde \rho = \rho/(2 \pi \ell_B
\sigma^2)$.
 
{\bf Attractive images}.
When $\Delta<0$, 
a small hard core of radius $b$ is required to prevent the thermodynamic
collapse of charges onto their images at the surface of the wall.
Irrespective of this steric regularization, 
the ground state is the same as when $\Delta=0$. It 
corresponds to the usual hexagonal Wigner layer of counter-ions
with the lattice spacing $a$ given by the electroneutrality
condition $q/\sigma = \sqrt{3}a^2/2$, localized at the hard-core distance 
$b$ from the wall \cite{Boroud05}.
We shift particles from their ground-state Wigner positions by small
displacements $z_j$. 
Expanding the two-body potential in (\ref{eq:energypi}) up to $z^2$ harmonic particle shifts,  
forgetting irrelevant constant terms, and keeping in 
mind that $\tilde{z}\ge \tilde{b}$ everywhere,
the energy change $\delta E$ from 
the ground state is given by
\begin{equation}
-\beta \delta E = \sum_{j=1}^N \left[ - (1+\Delta)\tilde{z}_j 
+ \frac{\Xi\vert\Delta\vert}{4\tilde{z}_j} \right] + S_z ,
\hbox{ where } S_z = \frac{\alpha^3}{4\sqrt{\Xi}} \sum_{j,k=1\atop (j\ne k)}^N
\frac{(\tilde{z}_j-\tilde{z}_k)^2
+ \Delta (\tilde{z}_j+\tilde{z}_k)^2}{(R_{jk}/a)^3}.
\label{eq:betaE}
\end{equation}
Here, all particles are indexed by an integer, and $R_{jk}$ denotes
the distance between ions $j$ and $k$ in the ground state structure.
In the remainder, we will refer to the one body term on the r.h.s. of
the first equation in
(\ref{eq:betaE}), corresponding to the square brackets, as $-\beta\delta  E_0$.
In other words, $-\beta\delta E = -\beta\delta E_0 +S_z$.
The associated one-body Boltzmann factor is dominant
at large $\Xi$, and reads
$\exp[-(1+\Delta)\tilde{z}+\Xi\vert\Delta\vert/(4\tilde{z})]$.
Up to a normalization factor, 
it therefore provides the leading order term in the 
counter-ion density expansion, that we denote $\tilde{\rho}_0(\tilde{z})$.
As expected with attractive images, this contribution stems 
from the region of $\tilde{z}$ close to $\tilde{b}$.

In order to compute the first sub-leading correction to the density
profile, we add to the one body potential $\delta E_0$ a generating
potential $u({\bf r})$, to be set to 0 at the end of the calculation.  We
next resort to the cumulant technique sketched in
\cite{Samaj11a}, where, having introduced $w({\bf r})=\exp(-\beta u({\bf r}))$ 
the Boltzmann weight of the generating potential,
the ionic density can be written as
\begin{equation}
\rho(z) = \rho_0(z) + \frac{\delta}{\delta w({\bf r})}\langle S_z \rangle_0
\Big\vert_{w({\bf r})=1} + \ldots
\end{equation} 
In this relation, the dots are for higher order
corrections, i.e. terms of larger power in $1/\Xi$, and
the statistical average $\langle ... \rangle_0$ refers to the system
of non-interacting particles with Boltzmann weight $\exp(-\beta \delta E_0)$.
Within this formalism, in the first correction order we have
\begin{equation}
\langle S_z \rangle_0 = \frac{3^{3/4}}{16 \pi^{3/2}\sqrt{\Xi}} N C_3
\left\{ (1+\Delta) [\tilde{z}^2]_0 - (1-\Delta) [\tilde{z}]_0^2 \right\} ,   
\end{equation} 
where
\begin{equation}
[\tilde{z}^p]_0 = \frac{\int_{z>b}d{\bf r}w({\bf r}) 
e^{-(1+\Delta)\tilde{z}+\Xi\vert\Delta\vert/(4\tilde{z})} \tilde{z}^p}{
\int_{z>b}d{\bf r}w({\bf r}) 
e^{-(1+\Delta)\tilde{z}+\Xi\vert\Delta\vert/(4\tilde{z})}} 
\quad \hbox{and} \quad
C_3 = \sum_{j,k=-\infty \atop (j,k)\neq (0,0)}^{\infty} \frac{1}{(j^2+jk+k^2)^{3/2}} \simeq 11.034.
\end{equation}
The density profile
then takes the form
\begin{equation} \label{eq:thedensityprofile}
\tilde{\rho}(\tilde{z})  \sim  
\frac{e^{-(1+\Delta)\tilde{z}+\Xi\vert\Delta\vert/(4\tilde{z})}}{ 
\int_{\tilde{b}}^{\infty} d\tilde{z} 
e^{-(1+\Delta)\tilde{z}+\Xi\vert\Delta\vert/(4\tilde{z})}} \Bigg\{ 1 +
\frac{3^{3/4}}{8\pi^{3/2}}\frac{C_3}{\sqrt{\Xi}} 
 \left[ \frac{1}{2}(1+\Delta)(\tilde{z}^2-a_2)
- (1-\Delta) a_1 (\tilde{z}-a_1) \right] \Bigg\} ,
\end{equation}
where
\begin{equation}
a_p = \frac{\int_{\tilde{b}}^{\infty} d\tilde{z}
e^{-(1+\Delta)\tilde{z}+\Xi\vert\Delta\vert/(4\tilde{z})} \tilde{z}^p}{
\int_{\tilde{b}}^{\infty}d\tilde{z} 
e^{-(1+\Delta)\tilde{z}+\Xi\vert\Delta\vert/(4\tilde{z})}} .
\end{equation}
This density profile satisfies the electroneutrality condition
 $\int \rho\, dz=\sigma/q$, which translates into 
 $\int\tilde\rho \, d\tilde z = 1$.
For the homogeneous dielectric case $\Delta=0$ and hard core $b=0$, 
we recover the result obtained in \cite{Samaj11a}.
For strictly negative values of $\Delta$, the effect of attractive
charge images is fundamental in the SC limit.   

{\bf Repulsive images}.
When $\Delta>0$ (repulsive image charges), the counter-ions are
on the one hand attracted to the wall by the surface charge $\sigma e$
and on the other hand repelled from the wall by their images.
It is natural to assume that the ground state of the system is formed by
the standard hexagonal Wigner crystal of counter-ions with lattice spacing 
$a$ given as previously by the electroneutrality condition
$q/\sigma = \sqrt{3} a^2/2$, localized at some distance $l$ from the wall.
This distance is determined by balancing the above 
attractive and repulsive forces.

The energy per particle for the Wigner crystal at distance $z$ from the wall 
is given by
\begin{equation}
\frac{\beta E_0(z)}{N}  = (1+\Delta)\frac{z}{\mu} + \frac{q^2 \ell_{\rm B}}{2}
\sum_{j\ne 1} \frac{1}{R_{1j}} 
 + \frac{q^2 \ell_{\rm B}}{2} \Delta \left[ \frac{1}{2z} + 
\sum_{j\ne 1} \frac{1}{\sqrt{R_{1j}^2+(2z)^2}} \right] .
\end{equation} 
On the rhs, the first term describes the interaction of the
particle with the surface charge, the second ($z$-in\-de\-pen\-dent) one
the direct interaction with all other particles, and the last two terms 
the interaction of the particle with its own self-image and with
images of all other particles. 
The distance of the Wigner crystal from the wall $l$ is determined by 
the stationarity condition
$\partial_z [\beta E_0(z)/N]\vert_{z=l}=0$.
Introducing the variable $t=2l/a$, this requirement can be written as
\begin{equation} \label{eq:deft}
\sum_{j,k=-\infty}^{\infty} \frac{1}{[t^2+(j^2+jk+k^2)]^{3/2}}
= \frac{4\pi}{\sqrt{3}} \frac{1+\Delta}{\Delta}\frac{1}{t} ,
\end{equation}
where use was made of the hexagonal structure.
The lattice sum appearing here can be represented in a more convenient way
for numerical calculation.
If $j+k$ is an even number, we introduce integers $n=(j+k)/2$ and $m=(j-k)/2$
which ``diagonalize'' the expression $j^2+jk+k^2 =3n^2 + m^2$.
If $j+k$ is an odd number, we introduce integers $n=(j+k+1)/2$ and 
$m=(j-k+1)/2$ which diagonalize $j^2+jk+k^2 =3(n-1/2)^2 + (m-1/2)^2$.
Finally, using the formula
~$
\Gamma(s) [f(n,m)]^{-s} =  \int_0^{\infty} du u^{s-1}
e^{-uf(n,m)} 
$,
we find that
\begin{equation}
\!\!\!\!\!\!\!\!\!\!\!\!\!\!\!\!\!\!\!\!\!\!\!\!\!
\sum_{j,k=-\infty}^{\infty} \frac{1}{[t^2+(j^2+jk+k^2)]^{3/2}} =
\frac{2}{\sqrt{\pi}} \int_0^{\infty} du \sqrt{u} e^{-u t^2} 
\left[
\theta_3(0,e^{-u}) \theta_3(0,e^{-3u}) + \theta_2(0,e^{-u}) \theta_2(0,e^{-3u})
\right] , 
\end{equation}
where
$
\theta_2(0,q) = \sum_n q^{(n-1/2)^2}$ and
$
\theta_3(0,q) = \sum_n q^{n^2} 
$ 
are the Jacobi theta functions \cite{Gradshteyn}, the sums running from
$-\infty$ to $\infty$.
The integral over theta functions can be evaluated with a high precision
and very quickly, e.g. by using the symbolic language {\it Mathematica}.
The numerical solution of Eq. (\ref {eq:deft}) for the extreme case 
$\Delta=1$ is $t\simeq 0.295$.
In the limit of small $\Delta$, we have $t\sim 3^{1/4}\sqrt{\Delta/(4\pi)}$.
For a non-negligible dielectric jump, with $\Delta$ of order unity,
the distance of the Wigner crystal from the wall $l=a t/2$
is of the order of the Wigner lattice spacing $a$.

We still need to examine the ground-state stability of the above Wigner 
crystal along all spatial directions.
The stability in the $(x,y)$ plane is related to
the already made normal mode analysis in Ref. \cite{Bonsall77}.
To study the stability of the Wigner crystal along the $z$ direction,
we shift one particle (say $j=1$) from its lattice position by a small
$\delta z = z-l$.
The corresponding change in the energy $\delta E$ is given by
\begin{eqnarray}
\beta \delta E & = & (1+\Delta)\frac{\delta z}{\mu} 
+ q^2 \ell_{\rm B} \sum_{j\ne 1}
\left[ \frac{1}{\sqrt{R_{1j}^2+\delta z^2}} - \frac{1}{R_{1j}} \right] 
+ \frac{q^2\ell_{\rm B}\Delta}{4} 
\left( \frac{1}{l+\delta z} - \frac{1}{l} \right) 
\nonumber \\ & & + \, q^2 \ell_{\rm B} \Delta \sum_{j\ne 1} \left[ 
\frac{1}{\sqrt{R_{1j}^2+(2l+\delta z)^2}} - 
\frac{1}{\sqrt{R_{1j}^2+(2l)^2}} \right] .
\nonumber \\ & &  
\end{eqnarray}
Expanding this expression in $\delta z$, the linear term vanishes due
to the stationarity condition (\ref{eq:deft}).
The quadratic term reads as $\beta\delta E = (\delta z)^2/\xi^2$,
where $\xi$ is given by
\begin{equation} \label{eq:defxi}
\left( \frac{a}{\xi} \right)^2  =  \frac{q^2\ell_{\rm B}}{a} \Bigg\{
\frac{\Delta}{t^3} - \frac{C_3}{2} - \frac{2\pi}{\sqrt{3}} \frac{1}{t}
 + \frac{3\Delta}{2} t^2 \sum_{j,k=-\infty}^{\infty} 
\frac{1}{[t^2+(j^2+jk+k^2)]^{5/2}} \Bigg\} .
\end{equation}
The criterion for the thermodynamic stability of the Wigner structure is 
that the sign of the contributions in curly brackets be positive, and 
it is so in the whole interval $\Delta\in [0,1]$.
For large but finite $\Xi$, the leading particle density profile is 
represented by the Gaussian distribution 
\begin{equation}
\rho(z) = \frac{\sigma}{q} \frac{e^{-(z-l)^2/\xi^2}}{
\int_0^{\infty} dz' e^{-(z'-l)^2/\xi^2}} .
\label{eq:gauss}
\end{equation}
Since from (\ref{eq:defxi}) the standard deviation $\xi$ scales like
$\xi/a\propto 1/\Xi^{1/4}$, in the limit $\Xi\to\infty$ the Gaussian 
distribution becomes the $\delta$-function peaked at $z=l$ and 
the particles are constrained to the ground-state Wigner lattice as 
it should be.

{\bf Conclusion}.
In this work, we have obtained the ionic profiles of counter-ions
strongly coupled to a planar charged interface, that defines 
the boundary of a large colloid which has a static dielectric constant
differing from that of the solvent (see Fig. \ref{fig:1}).
At variance with the Poisson-Boltzmann picture where dielectric images do
not affect the profiles, the strong-coupling (large $\Xi$) limit
is highly sensitive to the dielectric jump 
$\Delta = (\varepsilon-\varepsilon')/(\varepsilon+\varepsilon') \neq0$. 
When image charges are attractive ($\Delta<0$), a situation which requires
the introduction of an ionic hard-core to avoid instabilities,
the density profile is expectedly enhanced in the immediate
vicinity of the wall (see Eq. (\ref{eq:thedensityprofile})).
On the other hand, with repulsive images ($\Delta>0$), the 
ground state of the system is made up of an hexagonal Wigner
crystal at finite distance from the wall. Finite but large values of $\Xi$
turn the ground state configuration of counter-ions
into a Gaussian form with standard deviation proportional to
$a/\Xi^{1/4}$, where $a$ is the lattice spacing in the ground state Wigner
crystal. It would be desirable to test quantitatively our predictions
against numerical simulations, since those reported in a recent
analysis \cite{BdSL11}, in addition to pertaining to a curved geometry, 
also make use of a ion/wall hard core that prevents the study of the 
Gaussian behaviour embodied in Eq. (\ref{eq:gauss}). The two plates 
configuration is also an interesting venue for future work.

\begin{acknowledgement}
We would like to thank C. Texier for useful discussions.
The support received from the grants VEGA No. 2/0113/2009 and CE-SAS QUTE
is acknowledged. 
\end{acknowledgement}

\end{document}